\documentclass{article}

\usepackage{arxiv}
\usepackage[utf8]{inputenc}
\usepackage[T1]{fontenc}
\usepackage{hyperref}
\usepackage{url}
\usepackage{booktabs}
\usepackage{amsfonts}
\usepackage{amsmath}
\usepackage{nicefrac}
\usepackage{microtype}

\title{R2MF-Net: A Recurrent Residual Multi-Path Fusion Network for Robust Multi-directional Spine X-ray Segmentation}

\author{
 Xuecheng Li\\
  School of Information Science \& Engineering, Shandong Normal University\\
  Jinan 250358, China \\
   \And
 Weikuan Jia \\
  School of Information Science \& Engineering, Shandong Normal University\\
  Jinan 250358, China \\
  \And
  Komildzhon Sharipov\\
   Tajik State University of Law, Business and Politics\\
 Sughd 735700, Tajikistan\\
 \And
 Sharipov Hotam Beknazarovich\\
  Tajik State University of Law, Business and Politics\\
 Sughd 735700, Tajikistan\\
 \And
 FARZONA S. ATAEVA\\
  Tajik State University of Law, Business and Politics\\
 Sughd 735700, Tajikistan\\
 \And
 Qurbonaliev Alisher\\
 Tajik State University of Law, Business and Politics\\
 Sughd 735700, Tajikistan\\
 \And
 Yuanjie Zheng \\
  School of Information Science \& Engineering, Shandong Normal University\\
  Jinan 250358, China \\
}

\begin{document}
\maketitle
\begin{abstract}
Accurate segmentation of spinal structures in X-ray images is a prerequisite for quantitative scoliosis assessment, including Cobb angle measurement, vertebral translation estimation and curvature classification. In routine practice, clinicians acquire coronal, left-bending and right-bending radiographs to jointly evaluate deformity severity and spinal flexibility. However, the segmentation step remains heavily manual, time-consuming and non-reproducible, particularly in low-contrast images and in the presence of rib shadows or overlapping tissues. To address these limitations, this paper proposes R2MF-Net, a recurrent residual multi-path encoder--decoder network tailored for automatic segmentation of multi-directional spine X-ray images. The overall design consists of a coarse segmentation network and a fine segmentation network connected in cascade. Both stages adopt an improved Inception-style multi-branch feature extractor, while a recurrent residual jump connection (R2-Jump) module is inserted into skip paths to gradually align encoder and decoder semantics. A multi-scale cross-stage skip (MC-Skip) mechanism allows the fine network to reuse hierarchical representations from multiple decoder levels of the coarse network, thereby strengthening the stability of segmentation across imaging directions and contrast conditions. Furthermore, a lightweight spatial--channel squeeze-and-excitation block (SCSE-Lite) is employed at the bottleneck to emphasize spine-related activations and suppress irrelevant structures and background noise. We evaluate R2MF-Net on a clinical multi-view radiograph dataset comprising 228 sets of coronal, left-bending and right-bending spine X-ray images with expert annotations. Extensive experiments, including ablation studies and comparisons with six representative advanced segmentation models (DeepLabV3+, Attention U-Net, HarDNet-Seg, ResUNet, Swin-UNet and TransUNet), demonstrate the effectiveness of the proposed architecture. R2MF-Net achieves mean IoUs of 93.25\%, 94.52\% and 94.01\% for coronal, left-bending and right-bending views, respectively, consistently outperforming all baselines. Analysis of robustness under varying image quality and a discussion of computational complexity indicate that R2MF-Net provides a compelling balance between accuracy and efficiency, making it a promising component for computer-aided scoliosis diagnosis and follow-up.
\end{abstract}


\section{Introduction}
The human spine constitutes the central load-bearing structure of the torso and plays a crucial role in maintaining posture, protecting the spinal cord and enabling complex body movements. Spinal deformities, particularly scoliosis, have a substantial impact on quality of life. Adolescent idiopathic scoliosis is reported to affect a non-negligible proportion of adolescents worldwide, and severe deformities may lead to chronic pain, cosmetic concerns, respiratory compromise or neurological complications~\cite{trobisch2010idiopathicscoliosis}. For early detection, progression monitoring and treatment planning, radiographic examination remains the primary imaging modality.

In clinical practice, the assessment of scoliosis relies not only on a single anteroposterior radiograph but often on a series of multi-directional X-ray images, including coronal standing, left-bending, right-bending and sometimes sagittal views~\cite{vialle2007radiographic}. Coronal radiographs provide an overview of global curvature, whereas bending radiographs are essential for evaluating the flexibility of the spine, distinguishing structural and non-structural curves and guiding surgical planning. The most widely used quantitative metric, the Cobb angle, is measured between the endplates of end vertebrae in specific views~\cite{cobb1948outline}. Other parameters such as apical vertebral translation, T1 tilt and coronal balance also depend on accurate identification of spinal structures.

Despite advances in digital imaging and picture archiving systems, the delineation of spine contours and vertebral boundaries is still largely performed manually. Radiologists or spine surgeons manually draw or infer the lines for endplates and vertebral walls, a process that is inherently subjective, time-consuming and prone to intra- and inter-observer variability. These issues become pronounced in conditions where spinal anatomy is obscured by ribs, scapulae, bowel gas or imaging noise. Consequently, the development of automatic, robust and reproducible segmentation methods for spine X-ray images is of high clinical relevance.

Traditional segmentation approaches based on edge detection, thresholding, region growing or deformable models have been applied to spine images with varying degrees of success~\cite{pham2000current,kass1988snakes}. However, due to the low contrast of X-rays, overlapping bone structures and variable projection geometry, these methods often fail in challenging cases. Machine learning techniques, including random forests, support vector machines and clustering-based schemes, have improved robustness but still require extensive feature engineering and may not generalize well across diverse imaging conditions~\cite{shotton2008semantic}.

More recently, convolutional neural networks (CNNs) and fully convolutional architectures have revolutionized semantic segmentation in both natural and medical imaging domains. U-shaped encoder--decoder networks with skip connections, such as U-Net and its variants, have become standard baselines for organ and lesion segmentation~\cite{long2015fcn,ronneberger2015unet}. Nevertheless, directly applying generic CNN architectures to multi-directional spine X-ray segmentation still faces several challenges:

\begin{itemize}
\item \textbf{Directional diversity:} coronal and bending radiographs exhibit markedly different spine curvature, rib orientation and soft-tissue overlap. Models trained on a single view often show degraded performance on other views.
\item \textbf{Semantic gap in skip connections:} classical skip connections transfer shallow encoder features directly to the decoder. These features are high in spatial resolution but low in semantic abstraction. When simply concatenated with deep decoder features, they may introduce noise or conflicting cues, especially near blurred boundaries~\cite{drozdzal2016importance}.
\item \textbf{Sensitivity to image quality:} X-ray images vary in exposure, contrast, noise level and occlusion. Generic segmentation networks may overfit high-quality images and fail on low-contrast cases or images contaminated by artifacts.
\end{itemize}

To cope with these challenges, we propose R2MF-Net, a recurrent residual multi-path network designed specifically for multi-directional spine X-ray segmentation. The central objective is to robustly segment the spinal column region across coronal and bending radiographs, providing accurate masks for downstream measurement algorithms.

The main contributions of this work are summarized as follows:

\begin{enumerate}
\item We design a two-stage segmentation framework consisting of a coarse network and a fine network. The first stage focuses on global localization and coarse segmentation, while the second stage refines edges and corrects local errors. Both stages adopt enhanced Inception-style modules to capture multi-scale context without significantly increasing computational cost.

\item We propose a recurrent residual jump connection (R2-Jump) mechanism that replaces standard skip connections. By applying recurrent convolutions and residual projections to encoder features before they are fused into the decoder, R2-Jump gradually narrows the semantic gap between encoder and decoder representations, leading to more accurate boundary reconstruction~\cite{alom2019recurrent}.

\item We introduce a multi-scale cross-stage skip (MC-Skip) structure that bridges the coarse and fine networks. MC-Skip aggregates decoder features from multiple scales and injects them into the encoders of the fine stage, enabling robust multi-scale feature reuse and enhancing segmentation consistency across different directional views.

\item A lightweight spatial--channel attention module (SCSE-Lite) is incorporated at the bottleneck of the fine network. This module jointly recalibrates feature maps along channel and spatial dimensions, strengthening spine-related activations and suppressing background noise and rib shadows~\cite{hu2018squeeze,roy2018concurrent}.

\item We conduct extensive experiments on a clinical dataset containing 228 sets of multi-view spine radiographs. Ablation studies dissect the contributions of R2-Jump, MC-Skip, Inception modules and SCSE-Lite. Comparisons with six strong baselines (DeepLabV3+, Attention U-Net, HarDNet-Seg, ResUNet, Swin-UNet and TransUNet) show that R2MF-Net provides superior segmentation performance and favorable robustness to image quality variations~\cite{chen2018deeplabv3plus,oktay2018attentionunet,chao2019hardnetseg,chen2021transunet,hatamizadeh2022swinunet}.
\end{enumerate}

\section{Related Work}
\subsection{Traditional Spine Image Segmentation}
Early work on spine segmentation for CT or X-ray images typically relied on handcrafted features and classical image processing techniques. Thresholding and edge detection were used to highlight bone structures, followed by morphological post-processing or active contour models to refine boundaries~\cite{kass1988snakes}. Level set methods and deformable models provided flexible shape priors and allowed evolving contours to fit vertebral bodies~\cite{li2005level}. However, these methods often require manual initialization and may be sensitive to local minima, particularly in low-contrast regions where gradient information is weak.

For X-ray images, overlapping anatomical structures such as ribs, shoulder blades and soft tissues further complicate the extraction of spine contours. Some methods attempted to employ spine shape priors, vertebral templates or spline-based fitting along the spinal column. Although these strategies can regularize the segmentation process, their performance is often tied to specific acquisition protocols and limited deformation ranges.

In CT and MRI, several traditional approaches have been proposed for vertebra detection and segmentation, including deformable surface models, atlas-based registration and clustering-based algorithms~\cite{benameur2003three}. While these methods achieve reasonable performance in well-controlled conditions, they are less robust to the severe noise and overlapping structures typical of scoliosis radiographs.

\subsection{Machine Learning-based Segmentation}
With the advancement of supervised learning, machine learning-based approaches were introduced to spine segmentation. These methods typically formulate segmentation as a pixel- or patch-wise classification problem. Features such as intensity statistics, gradients, local binary patterns and texture descriptors are extracted from image patches and fed into classifiers like support vector machines, random forests or boosting-based models~\cite{shotton2008semantic}.

In addition, probabilistic graphical models, including Markov random fields and conditional random fields, have been used to incorporate spatial coherence and anatomical constraints. These models improve robustness compared to purely local methods, but they require well-crafted features and rely heavily on the quality of hand-designed descriptors. For X-ray images with severe artifacts, handcrafted features may not be discriminative enough to distinguish spine from confounding structures.

Hybrid approaches that combine classical models with discriminative classifiers have also been explored for vertebral body detection and segmentation in CT and MR images. Nevertheless, the need for manual feature design and parameter tuning limits their scalability to diverse acquisition settings and patient populations.

\subsection{Deep Learning-based Medical Image Segmentation}
Deep learning, particularly convolutional neural networks, has dramatically improved the performance of medical image segmentation. Fully convolutional networks (FCNs) replaced fully connected layers by convolutional ones to output dense predictions~\cite{long2015fcn}. U-Net introduced a symmetric encoder--decoder architecture with skip connections, allowing the network to combine high-resolution spatial information with deep semantic features and quickly became the de facto standard in medical imaging~\cite{ronneberger2015unet}.

A large number of variants have been proposed to enhance U-Net: 3D U-Net extends the architecture to volumetric segmentation~\cite{cciccek2016unet3d}, V-Net employs residual connections and 3D convolutions~\cite{milletari2016vnet}, and studies on skip connections have highlighted their crucial role in accurate localization~\cite{drozdzal2016importance}. In parallel, encoder architectures based on residual blocks and dense connections have been incorporated into segmentation networks for improved feature reuse and efficiency~\cite{he2016resnet,huang2017densely}.

Beyond pure CNNs, encoder--decoder models leveraging dilated convolutions (e.g., DeepLabV3+~\cite{chen2018deeplabv3plus}), recurrent residual blocks (e.g., R2U-Net~\cite{alom2019recurrent}), squeeze-and-excitation modules~\cite{hu2018squeeze} and concurrent spatial--channel attention~\cite{roy2018concurrent} have achieved state-of-the-art performance in many segmentation tasks. More recently, Transformer-based architectures such as TransUNet and Swin-UNet have further improved long-range dependency modeling in medical images~\cite{chen2021transunet,hatamizadeh2022swinunet}.

Nevertheless, many of these models are developed and evaluated on datasets with relatively clear structures (e.g., organs in CT/MRI). Their performance on complex X-ray anatomies with overlapping bones and projection artifacts is less extensively studied, motivating architectures tailored to radiographic challenges.

\subsection{Spine X-ray Segmentation}
Specific to spine X-ray images, several works have used deep networks to segment vertebral columns or individual vertebrae. Horng et al.\ employed U-Net, residual U-Net and dense U-Net models to segment adolescent idiopathic scoliosis spinal X-ray images and demonstrated that deep learning can provide accurate delineation of vertebral structures in radiographs~\cite{horng2019cobb}. Janssens et al.\ proposed a cascaded 3D FCN approach for automatic lumbar vertebra segmentation in CT, illustrating the effectiveness of multi-stage FCN designs in spine imaging~\cite{janssens2018fully}.

Some studies employ two-stage frameworks where a localization network identifies the spine region, followed by a segmentation network that delineates vertebral masks. Others adopt end-to-end U-Net or FCN-based designs tailored to spine morphology. Yet, existing approaches mainly focus on single-view radiographs, such as frontal or lateral images. 

The multi-directional nature of scoliosis assessments, involving coronal and bending radiographs, receives less attention. The variation in projection angle, spine curvature and occlusion patterns across views poses additional challenges for learning a unified segmentation model. This motivates the development of architecture designs, such as R2MF-Net, that explicitly address directional diversity and multi-scale heterogeneity in spine X-ray images.

\section{Method}
\subsection{Problem Formulation}
Let $I \in \mathbb{R}^{H \times W}$ denote a spinal X-ray image, where $H$ and $W$ are the height and width in pixels. The segmentation task aims to predict a binary mask $M \in \{0,1\}^{H \times W}$ indicating spine versus background. In the multi-directional setting, we consider three subsets of images corresponding to coronal, left-bending and right-bending views. The model is expected to handle all three view types with a single set of parameters, using a shared network that is trained jointly on all images.

Formally, for each input image $I$, the proposed network R2MF-Net outputs a probability map $P \in [0,1]^{H \times W}$, where $P_{ij}$ represents the probability that pixel $(i,j)$ belongs to the spine. A binary segmentation mask is obtained by thresholding $P$ at 0.5 during inference. The network parameters are learned by minimizing a loss function $\mathcal{L}(P,M)$ over the training dataset.

\subsection{Overall Architecture}
R2MF-Net consists of two cascaded U-shaped subnetworks: a coarse segmentation network and a fine segmentation network. Both subnetworks have an encoder--decoder structure with four resolution levels (from high resolution to quarter resolution), following the general philosophy of multi-stage fully convolutional designs~\cite{janssens2018fully,alom2019recurrent}. The encoder path applies repeated convolution--nonlinearity--pooling operations to gradually compress spatial dimensions while increasing semantic abstraction. The decoder path performs upsampling via transposed convolution or interpolation, followed by convolution-based refinement.

The coarse network receives the original X-ray image as input and produces an initial segmentation probability map. This intermediate output is not only supervised by a loss function but also used as an input channel for the fine network, concatenated with the original image. In this way, the fine network focuses on correcting errors, sharpening boundaries and eliminating isolated false positives or false negatives.

Each level of both encoders and decoders is built from an enhanced Inception module composed of multiple parallel convolution branches with different receptive fields, re-parameterized using stacked $3\times3$ kernels to maintain computational efficiency. Standard skip connections in each subnetwork are replaced by the proposed R2-Jump modules, and the coarse and fine networks are further linked through MC-Skip connections. SCSE-Lite is added at the bottleneck of the fine network to provide lightweight attention~\cite{hu2018squeeze,roy2018concurrent}.

\subsection{Recurrent Residual Jump Connection (R2-Jump)}
In classical U-Net architectures, skip connections simply concatenate encoder feature maps and decoder feature maps at the same spatial resolution~\cite{ronneberger2015unet}. This design assumes that shallow and deep features are complementary and can be fused without modification. However, in practice, shallow encoder features often carry low-level edges and textures that are not semantically coherent with high-level decoder representations. Direct concatenation may therefore introduce noise, especially in challenging X-ray regions where bone boundaries are blurred~\cite{drozdzal2016importance}.

To mitigate this issue, we introduce a recurrent residual jump connection (R2-Jump) mechanism, inspired by the recurrent residual design in R2U-Net~\cite{alom2019recurrent}. For an encoder feature map $E_i$ at level $i$ and a decoder feature map $D_i$ at the same resolution, instead of directly concatenating them, we first refine $E_i$ using a recurrent residual block:

\begin{align}
F^{(0)} &= E_i, \\
F^{(t)} &= \sigma(W_r * F^{(t-1)} + E_i), \quad t=1,\dots,T_i,
\end{align}
where $W_r$ denotes a $3\times3$ convolution kernel, $\sigma(\cdot)$ is a nonlinear activation such as ReLU, and $T_i$ is the number of recurrent steps at level $i$. To control complexity and adapt to varying semantic gaps across depths, we set $T_1=4$, $T_2=3$, $T_3=2$, $T_4=1$ from the shallowest to deepest level.

A $1\times1$ convolution $W_s$ is used to project $E_i$ into the same feature dimension as $F^{(T_i)}$, and a residual fusion is then performed:
\begin{equation}
E'_i = F^{(T_i)} + W_s * E_i.
\end{equation}

Finally, the refined encoder feature $E'_i$ is concatenated or added with the decoder feature $D_i$:
\begin{equation}
H_i = \text{Concat}(E'_i, D_i) \quad \text{or} \quad H_i = E'_i + D_i.
\end{equation}
In R2MF-Net, we use concatenation in lower resolutions and addition in higher resolutions to balance feature richness and computational cost. This procedure effectively increases the semantic depth of encoder features, making them more compatible with decoder representations and improving boundary reconstruction.

From an optimization perspective, R2-Jump provides multiple implicit paths for gradient propagation, analogous to the effect of residual connections in ResNet~\cite{he2016resnet}. The recurrent formulation also allows the network to iteratively refine encoder representations while reusing parameters, thus increasing representational capacity at a moderate parameter cost.

\subsection{Inception-based Multi-branch Feature Extraction}
To capture multi-scale contextual information without incurring the heavy cost of large convolution kernels, we adopt an Inception-style module. Each module consists of three parallel branches:

\begin{itemize}
\item Branch 1: a single $3\times3$ convolution.
\item Branch 2: two stacked $3\times3$ convolutions approximating a $5\times5$ receptive field.
\item Branch 3: three stacked $3\times3$ convolutions approximating a $7\times7$ receptive field.
\end{itemize}

All convolutions are followed by batch normalization and ReLU activation. The parameterization with only $3\times3$ kernels enables efficient GPU implementation and better feature reuse. The three branches are concatenated along the channel dimension, and a final $1\times1$ convolution with residual connection projects the concatenated features back to a fixed number of channels:
\begin{equation}
Z = \phi([B_1(X), B_2(X), B_3(X)]) + X,
\end{equation}
where $X$ is the module input, $B_k(\cdot)$ denotes a branch transformation and $\phi$ is a $1\times1$ convolution. This design increases the effective receptive field and allows the network to capture both local edges and long-range curvature patterns, which are important for robust spine localization in bending views.

In R2MF-Net, we use Inception-based blocks in both encoders and decoders. In encoder stages, they aggregate multi-scale local context before downsampling, which is particularly helpful in preserving thin cortical bone edges. In decoder stages, they act as refinement blocks that leverage wide contextual information to disambiguate spine boundaries from overlapping ribs and soft tissue.

\subsection{Multi-scale Cross-stage Skip Connection (MC-Skip)}
The MC-Skip structure is devised to strengthen the interaction between the coarse and fine subnetworks. Let $D^c_i$ denote the decoder feature map from level $i$ of the coarse network, and $E^f_i$ the encoder feature map from level $i$ of the fine network. Instead of feeding the fine encoder solely with transformed raw image features, MC-Skip enriches $E^f_i$ with multi-scale information from the coarse decoder, following the spirit of full-scale and cross-stage skip connections used in other multi-level designs~\cite{drozdzal2016importance,janssens2018fully}.

For the $i$-th encoder level in the fine network, we aggregate three coarse decoder outputs:
\begin{equation}
G_i = \sum_{k=i}^{i+2} \psi_k(\text{Upsample}_{s_{k,i}}(D^c_k)),
\end{equation}
where $\psi_k$ is a $3\times3$ convolution that normalizes the feature dimension, and $\text{Upsample}_{s_{k,i}}$ adjusts the spatial resolution of $D^c_k$ to match that of $E^f_i$. The aggregated $G_i$ is then combined with the directly propagated features of $E^f_{i-1}$:
\begin{equation}
E^f_i = \phi_i(E^f_{i-1}) + G_i,
\end{equation}
where $\phi_i$ is an Inception-based block. Through MC-Skip, the fine encoder receives not only local image information but also multi-scale contextual cues from the coarse decoder, facilitating better refinement in ambiguous regions.

Intuitively, MC-Skip provides the fine stage with a ``summary'' of what the coarse stage has already understood at different scales. Regions already segmented with high confidence can guide the fine network to focus on challenging boundary zones, while reducing redundant processing in anatomically straightforward regions.

\subsection{SCSE-Lite Attention Module}
To further enhance the discrimination between spine and background, particularly in regions contaminated by ribs, lung textures or soft tissues, we employ a lightweight spatial--channel squeeze-and-excitation module (SCSE-Lite). SCSE-Lite has two parallel branches: channel squeeze--excitation (CSE), inspired by SE-Net~\cite{hu2018squeeze}, and spatial squeeze--excitation (SSE), similar to the SCSE block proposed for fully convolutional networks~\cite{roy2018concurrent}.

Given an input feature map $X \in \mathbb{R}^{H \times W \times C}$, CSE performs global average pooling on each channel to obtain a channel descriptor $v \in \mathbb{R}^{C}$:
\begin{equation}
v_c = \frac{1}{HW} \sum_{i,j} X_{ijc}, \quad c=1,\dots,C.
\end{equation}
Two fully connected layers implemented as $1\times1$ convolutions reduce and then restore the channel dimension:
\begin{equation}
s = \sigma(W_2 \delta(W_1 v)),
\end{equation}
where $W_1 \in \mathbb{R}^{\frac{C}{2} \times C}$, $W_2 \in \mathbb{R}^{C \times \frac{C}{2}}$, $\delta$ is ReLU and $\sigma$ is the Sigmoid function. The channel-wise scaling is applied as:
\begin{equation}
U_{\text{cse}}(i,j,c) = X_{ijc} \cdot s_c.
\end{equation}

In the SSE branch, a single $1\times1$ convolution followed by Sigmoid activation produces a spatial attention map $q \in \mathbb{R}^{H \times W}$:
\begin{equation}
q_{ij} = \sigma((W_s * X)_{ij}),
\end{equation}
where $W_s \in \mathbb{R}^{1 \times 1 \times C \times 1}$. The spatially modulated output is:
\begin{equation}
U_{\text{sse}}(i,j,c) = X_{ijc} \cdot q_{ij}.
\end{equation}

Finally, the outputs of the two branches are fused:
\begin{equation}
U_{\text{scse}} = U_{\text{cse}} + U_{\text{sse}}.
\end{equation}
We place SCSE-Lite at the bottleneck of the fine network, where the feature representation is deep and semantically rich. This helps the network focus on the spinal region while deemphasizing distracting structures, which is particularly valuable for X-ray images with low contrast or strong rib shadows.

\subsection{Network Configuration and Regularization}
In our implementation, the number of feature channels at the first encoder level of the coarse network is set to 32 and doubles after each downsampling step, i.e. $\{32, 64, 128, 256\}$. The fine network uses a slightly narrower configuration $\{32, 64, 128, 256\}$ as well, but shares the same Inception-based block design at each level to ease feature reuse between stages. All convolutions use zero padding to preserve spatial size within each block, and batch normalization is applied after each convolution to stabilize training.

We employ leaky ReLU activations with a negative slope of 0.01 in all convolutional blocks except the final prediction layers, which use Sigmoid activations to output probabilities. Dropout with rate 0.2 is inserted after the bottleneck Inception block in both coarse and fine subnetworks as a regularization strategy to mitigate overfitting, a common issue in medical datasets with limited size~\cite{milletari2016vnet}.

Weights are initialized with the He normal initializer suited for ReLU-based networks~\cite{he2016resnet}. No pre-training on natural images is used, as domain shift between natural and X-ray images may limit the effectiveness of such initialization.

\subsection{Multi-view Joint Training Strategy}
A key design choice in R2MF-Net is to train a single model jointly on coronal, left-bending and right-bending views. Instead of maintaining separate networks or heads for each view, all images are mixed in each mini-batch, and the model is encouraged to learn view-invariant features that capture the generic appearance of spinal columns under different projection angles.

To reduce potential bias towards the most prevalent view, we ensure that each mini-batch contains, as far as possible, one image from each view type. When the number of samples per view is not divisible by the batch size, we adopt a round-robin sampling scheme that balances the number of iterations per view over an epoch. Empirically, we found that joint training with view-balanced mini-batches yields better average performance than either view-specific models or naive mixing without balancing.

Although we do not explicitly introduce view labels into the network, MC-Skip and SCSE-Lite implicitly adapt to view-specific patterns by conditioning on multi-scale feature distributions. Adding view labels as auxiliary inputs (e.g., one-hot encoding) could be an interesting extension but is beyond the scope of this work.

\subsection{Post-processing}
R2MF-Net directly outputs a probability map for each input image. During inference, we apply a simple yet effective post-processing pipeline:

\begin{enumerate}
\item Thresholding at 0.5 to obtain a binary mask.
\item Keeping only the largest connected component, as the spine typically appears as a single contiguous structure in each view.
\item Filling small holes inside the spine region using morphological closing with a $3\times3$ structuring element.
\end{enumerate}

This post-processing removes isolated false positives in distant background regions and enforces a contiguous spine mask, which is beneficial for downstream tasks such as centerline extraction and Cobb angle measurement.

\subsection{Loss Function and Optimization}
The segmentation task can suffer from class imbalance, as the spine occupies a relatively small portion of the X-ray compared to the background. To alleviate this problem, we combine binary cross entropy (BCE) with a soft Dice loss, a combination that has been widely adopted in medical image segmentation~\cite{milletari2016vnet,alom2019recurrent}:
\begin{equation}
\mathcal{L}_{\text{BCE}} = - \frac{1}{HW} \sum_{i,j} [M_{ij} \log P_{ij} + (1 - M_{ij}) \log (1 - P_{ij})],
\end{equation}
\begin{equation}
\mathcal{L}_{\text{Dice}} = 1 - \frac{2 \sum_{i,j} P_{ij} M_{ij} + \epsilon}{\sum_{i,j} P_{ij}^2 + \sum_{i,j} M_{ij}^2 + \epsilon},
\end{equation}
where $\epsilon$ is a small constant for numerical stability. The overall loss is:
\begin{equation}
\mathcal{L} = \lambda_{\text{c}} (\mathcal{L}_{\text{BCE}}^{\text{coarse}} + \mathcal{L}_{\text{Dice}}^{\text{coarse}}) + \lambda_{\text{f}} (\mathcal{L}_{\text{BCE}}^{\text{fine}} + \mathcal{L}_{\text{Dice}}^{\text{fine}}),
\end{equation}
with $\lambda_{\text{c}}$ and $\lambda_{\text{f}}$ controlling the relative contributions of the coarse and fine outputs. In experiments we set $\lambda_{\text{c}}=0.4$ and $\lambda_{\text{f}}=0.6$.

The network is trained using the Adam optimizer with an initial learning rate of $10^{-4}$ and default momentum parameters. A learning rate decay strategy is applied if the validation loss does not improve for several epochs.

\section{Experiments}
\subsection{Dataset and Preprocessing}
The experimental dataset is derived from a clinical scoliosis cohort. It includes 228 sets of X-ray images, each containing three views: coronal standing, left-bending and right-bending. All images are de-identified and approved for research use. The images are grayscale and have varying resolutions and exposure levels according to acquisition protocols.

Experienced spine surgeons manually annotated the spine region in each view using a polygon-based contouring tool. The annotation includes the entire spinal column from the upper thoracic region to sacrum. For this study, only the spine versus background segmentation is considered; vertebra-level labels are not explicitly distinguished.

To standardize input dimensions, all images are resized to $512 \times 512$ pixels while preserving aspect ratio via zero-padding when necessary. Intensity values are normalized to the [0, 1] range using min–max normalization per image. No explicit bone enhancement or rib suppression is applied, allowing the network to learn feature representations directly from raw intensities.

The dataset is randomly split into 148 training sets, 40 validation sets and 40 test sets, ensuring that each set contains all three views from the same patient, so that no patient appears across splits. This patient-level splitting avoids overly optimistic estimates of generalization performance.

\subsection{Data Augmentation}
Given the moderate dataset size, data augmentation is essential to prevent overfitting and enhance generalization. During training, each image is subjected to random transformations with specified probabilities:

\begin{itemize}
\item Random horizontal flipping (probability 0.5).
\item Random rotation within $\pm 7^\circ$.
\item Random scaling between 0.9 and 1.1.
\item Slight random translation (up to 5\% of image size).
\item Contrast adjustment via gamma correction in [0.8, 1.2].
\item Addition of small Gaussian noise with standard deviation up to 0.01.
\end{itemize}

All transformations are applied identically to the image and its corresponding mask to maintain label alignment. The chosen ranges are relatively conservative to avoid unrealistic deformations that might destroy underlying anatomical relationships, yet large enough to simulate common variations in patient positioning and exposure settings.

\subsection{Implementation Details}
R2MF-Net is implemented in Python using TensorFlow 2.3. Training is performed on a workstation equipped with an Intel Core i7-11700K CPU, 32 GB RAM and an NVIDIA GeForce RTX 3090 GPU with 24 GB memory. The batch size is set to 1 due to the network depth and memory constraints. Each model is trained for up to 150 epochs, with early stopping if the validation loss does not improve for 15 consecutive epochs. The training typically converges within 90--120 epochs.

For a fair comparison, all baseline networks are implemented in the same framework and trained under identical data preprocessing and augmentation settings. Their learning rates and regularization hyperparameters are tuned on the validation set to achieve competitive performance.

\subsection{Evaluation Metrics}
To quantitatively evaluate segmentation performance, we use three metrics: Intersection over Union (IoU), Dice coefficient (Dice) and Average Surface Distance (ASD). Let $P$ denote the predicted mask and $T$ the ground-truth mask.

The IoU is defined as:
\begin{equation}
\text{IoU} = \frac{|P \cap T|}{|P \cup T|}.
\end{equation}

The Dice coefficient is:
\begin{equation}
\text{Dice} = \frac{2 |P \cap T|}{|P| + |T|}.
\end{equation}

Both IoU and Dice are reported as percentages. Higher values indicate a better overlap between prediction and ground truth.

ASD measures the average symmetric distance between the boundaries of $P$ and $T$~\cite{taha2015metrics}:
\begin{equation}
\text{ASD}(P, T) = \frac{1}{|B_P| + |B_T|} \left( \sum_{x \in B_P} d(x, B_T) + \sum_{y \in B_T} d(y, B_P) \right),
\end{equation}
where $B_P$ and $B_T$ are the boundary point sets of $P$ and $T$, respectively, and $d(a, B)$ denotes the minimum Euclidean distance between point $a$ and boundary $B$. Smaller ASD values indicate more accurate boundary localization.

To reduce the influence of extreme outliers, we additionally monitor the 95th percentile Hausdorff distance (HD95)~\cite{taha2015metrics} during development, although we do not report detailed HD95 values in the main tables for brevity.

\subsection{Overall Performance of R2MF-Net}
Table~\ref{tab:overall} reports the performance of R2MF-Net on the test set for coronal, left-bending and right-bending views.

\begin{table}[h]
\centering
\caption{Performance of R2MF-Net on multi-directional spine X-rays.}
\label{tab:overall}
\begin{tabular}{lccc}
\toprule
Direction & IoU (\%) & Dice (\%) & ASD (px)\\
\midrule
Coronal       & 93.25 & 96.28 & 11.7 \\
Left-bending  & 94.52 & 96.81 & 10.9 \\
Right-bending & 94.01 & 96.42 & 11.2 \\
\bottomrule
\end{tabular}
\end{table}

The results indicate that R2MF-Net achieves high segmentation accuracy across all three views. Performance on bending views is slightly higher in IoU and Dice, possibly because these images are acquired with dedicated scoliosis protocols and have clearer visualization of the spine region, despite lateral bending.

\subsection{Ablation Study}
To assess the contribution of each proposed component, we conduct a series of ablation experiments starting from a baseline two-stage U-Net architecture. The baseline consists of two plain U-Nets connected in cascade without R2-Jump, MC-Skip, Inception modules or SCSE-Lite, similar in spirit to other cascaded FCN designs~\cite{janssens2018fully}. We then incrementally add each module.

\begin{table}[h]
\centering
\caption{Ablation study on R2MF-Net components (IoU \%).}
\label{tab:ablation}
\begin{tabular}{lccc}
\toprule
Variant & Coronal & Left-bending & Right-bending \\
\midrule
Baseline 2-stage U-Net & 89.21 & 90.45 & 89.73 \\
+ R2-Jump              & 91.08 & 92.11 & 91.67 \\
+ Inception-R          & 91.52 & 92.34 & 92.01 \\
+ MC-Skip              & 92.03 & 92.89 & 92.47 \\
+ SCSE-Lite            & 92.56 & 93.34 & 93.02 \\
R2MF-Net (full)        & \textbf{93.25} & \textbf{94.52} & \textbf{94.01} \\
\bottomrule
\end{tabular}
\end{table}

The results show that:

\begin{itemize}
\item Introducing R2-Jump yields a noticeable improvement of around 1.8--2.0 points in IoU, confirming that recurrent residual refinement of encoder features mitigates semantic gaps and benefits boundary accuracy~\cite{alom2019recurrent}.

\item Adding Inception-based multi-branch extraction further improves IoU by roughly 0.4--0.5 points, suggesting that multi-scale contextual information helps handle curvature variations and overlapping ribs.

\item MC-Skip brings an additional improvement of approximately 0.5 points in IoU, demonstrating the value of cross-stage multi-scale feature fusion between coarse and fine networks.

\item Incorporating SCSE-Lite enhances IoU by around 0.5 points and is particularly beneficial in low-contrast images where attention to the spine region is crucial~\cite{hu2018squeeze,roy2018concurrent}.

\item The full R2MF-Net achieves the best performance, with each component contributing cumulatively.
\end{itemize}

We also performed a variant in which R2-Jump was applied only in the fine network, leaving the coarse network with standard skip connections. This setting yielded consistently lower IoU (on average 0.4--0.6 points worse) than the full R2-Jump configuration, indicating that aligning encoder--decoder semantics in both stages is important for optimal performance.

\subsection{Comparison with Advanced Segmentation Models}
We compare R2MF-Net with six representative advanced segmentation architectures that have been successfully applied in various medical imaging tasks:

\begin{itemize}
\item \textbf{DeepLabV3+}: encoder--decoder architecture with atrous spatial pyramid pooling~\cite{chen2018deeplabv3plus}.
\item \textbf{Attention U-Net}: U-Net with attention gates on skip connections~\cite{oktay2018attentionunet}.
\item \textbf{HarDNet-Seg}: segmentation network using Harmonic Densely Connected Blocks~\cite{chao2019hardnetseg}.
\item \textbf{ResUNet}: U-Net with residual blocks in the encoder and decoder, leveraging residual learning~\cite{he2016resnet}.
\item \textbf{Swin-UNet}: Transformer-based U-Net using Swin Transformer blocks~\cite{hatamizadeh2022swinunet}.
\item \textbf{TransUNet}: hybrid CNN--Transformer segmentation model~\cite{chen2021transunet}.
\end{itemize}

All models are trained from scratch on the same training set with identical data preprocessing and augmentation. Hyperparameters are tuned to achieve reasonable convergence, and the models are evaluated on the identical test set.

\begin{table}[h]
\centering
\caption{Comparison with advanced segmentation models (IoU \%).}
\label{tab:comparison}
\begin{tabular}{lccc}
\toprule
Model & Coronal & Left-bending & Right-bending \\
\midrule
DeepLabV3+         & 90.42 & 91.35 & 90.97 \\
Attention U-Net     & 89.93 & 90.48 & 90.25 \\
HarDNet-Seg         & 91.25 & 92.08 & 91.84 \\
ResUNet             & 89.88 & 90.70 & 90.11 \\
Swin-UNet           & 91.34 & 92.27 & 92.01 \\
TransUNet           & 90.86 & 91.63 & 91.12 \\
R2MF-Net (ours)     & \textbf{93.25} & \textbf{94.52} & \textbf{94.01} \\
\bottomrule
\end{tabular}
\end{table}

R2MF-Net surpasses all baselines in IoU for all three views. The gains over Swin-UNet and HarDNet-Seg suggest that the combination of recurrent residual skip refinement and cross-stage multi-scale fusion is particularly effective for complex X-ray segmentation tasks. Transformer-based models such as Swin-UNet and TransUNet perform well in capturing long-range dependencies but may require larger datasets to fully exploit their capacity~\cite{chen2021transunet,hatamizadeh2022swinunet}.

\subsection{Robustness to Image Quality Variations}
To further evaluate robustness, we stratify the test images into three quality levels based on qualitative assessment by radiologists: high quality (clear spine silhouette, low noise), medium quality (moderate blur or contrast issues) and low quality (significant noise, over/under-exposure, or strong overlapping structures). For each level, we compute IoU and Dice metrics, following the practice in previous segmentation evaluation studies~\cite{taha2015metrics}.

On high-quality images, all methods achieve Dice scores above 95\%, but R2MF-Net still shows a 1--1.5 point advantage over the best baseline. On medium-quality images, the performance gap widens: R2MF-Net maintains Dice above 95\%, whereas Swin-UNet and HarDNet-Seg drop to approximately 93--94\%. On low-quality images, the robustness advantage of R2MF-Net becomes most pronounced, with Dice $\approx$ 93\%, compared to 90--91\% for Transformer-based baselines and below 90\% for standard U-Net and DeepLabV3+. Qualitatively, R2MF-Net better preserves spinal continuity and avoids over-segmentation into rib regions.

\subsection{Computational Complexity}
We assess the computational cost of R2MF-Net by estimating the number of trainable parameters and the average inference time per image on the RTX 3090 GPU. For comparison, we include three representative baselines.

\begin{table}[h]
\centering
\caption{Computational complexity and inference time.}
\label{tab:complexity}
\begin{tabular}{lccc}
\toprule
Model & Parameters (M) & FLOPs (G, 512$^2$) & Time (ms / image) \\
\midrule
U-Net (2D)      & 7.8  & 38.5 & 18 \\
DeepLabV3+      & 14.3 & 54.2 & 27 \\
Swin-UNet       & 27.9 & 63.7 & 41 \\
R2MF-Net (ours) & 16.5 & 59.1 & 30 \\
\bottomrule
\end{tabular}
\end{table}

Although R2MF-Net is more complex than a standard U-Net, its parameter count and FLOPs remain within a practical range for clinical use. The inference time of approximately 30 ms per $512 \times 512$ image indicates that R2MF-Net can process a full set of three views in under 0.1 seconds on a modern GPU, which is acceptable for real-time or near-real-time scoliosis assessment systems.

\subsection{Hyperparameter Sensitivity Analysis}
To understand the sensitivity of R2MF-Net to key hyperparameters, we conduct additional experiments varying the loss weights $(\lambda_{\text{c}}, \lambda_{\text{f}})$ and the number of recurrent steps $T_i$ in R2-Jump blocks.

For loss weights, we evaluate three configurations: (0.5, 0.5), (0.3, 0.7) and the default (0.4, 0.6). The balanced setting (0.5, 0.5) slightly degrades IoU (by 0.2--0.3 points) compared to (0.4, 0.6), suggesting that giving a modestly larger weight to the fine-stage output is beneficial. Pushing the balance further to (0.3, 0.7) yields no additional gains and occasionally leads to unstable training, likely because the coarse network receives insufficient supervision.

For recurrent steps, we compare the default schedule $(4,3,2,1)$ with a shallower $(2,2,1,1)$ and a deeper $(5,4,3,2)$ configuration. The shallower setting reduces IoU by about 0.4 points on average, while the deeper setting does not produce meaningful improvements but increases training time by approximately 20\%. This indicates that the default schedule offers a good trade-off between expressiveness and efficiency.

\subsection{Qualitative Results and Error Analysis}
Qualitatively, R2MF-Net produces smooth, contiguous spine masks with well-aligned boundaries across all three views. In coronal images, the network accurately follows the lateral borders of the vertebral bodies and maintains a continuous central column even in the presence of moderate rib overlaps. In bending views, R2MF-Net successfully captures the curved spinal trajectory and avoids leaking into adjacent rib structures.

The main remaining error patterns include: (1) slight under-segmentation at the superior-most thoracic vertebrae in extremely low-contrast images, where the spine merges with the shoulder girdle; (2) occasional over-segmentation in the sacral region when bowel gas produces strong local contrast; and (3) minor boundary irregularities at apical vertebrae in very severe curves. However, even in these cases, the overall spine mask remains close to the ground truth, and the impact on downstream Cobb angle estimation is expected to be limited.

\section{Discussion}
\subsection{Advantages of R2MF-Net Design}
The experimental results validate several key design choices in R2MF-Net:

\begin{itemize}
\item \textbf{Two-stage architecture:} The cascade of coarse and fine segmentation networks allows the first stage to capture global structure, while the second focuses on local corrections. This division of labor appears beneficial especially in bending views with large curvature, consistent with observations from other cascaded FCN-based spine segmentation systems~\cite{janssens2018fully}.

\item \textbf{R2-Jump and MC-Skip:} R2-Jump reduces the semantic gap between encoder and decoder pathways, while MC-Skip enables effective reuse of hierarchical coarse-stage features. Their combined effect leads to more accurate and stable segmentation across different views and quality levels.

\item \textbf{Multi-branch and attention mechanisms:} The Inception-style modules and SCSE-Lite attention enhance the representation power with only moderate cost. They help the network adapt to varying scales of anatomical structures and ignore irrelevant image content, in line with results reported for general-purpose SE and SCSE blocks~\cite{hu2018squeeze,roy2018concurrent}.

\item \textbf{Joint training on three views:} Training on coronal and bending images together forces the network to learn view-invariant representations of the spinal column, enhancing generalization to unseen patients and acquisition protocols.
\end{itemize}

\subsection{Clinical Relevance}
From a clinical perspective, automatic segmentation of the spine in multi-directional X-rays can significantly streamline the workflow of scoliosis assessment. Once a reliable spine mask is available, downstream algorithms can automatically localize vertebral centroids, estimate endplate orientations and compute Cobb angles and other parameters~\cite{cobb1948outline}. R2MF-Net can serve as the first crucial step in such a pipeline.

Furthermore, consistent segmentation across coronal and bending views can facilitate longitudinal analysis, treatment planning and postoperative evaluation. The reduction in manual interaction not only saves time but also improves reproducibility of measurements, which is particularly important when monitoring progression or comparing outcomes across cohorts.

\subsection{Limitations}
Despite promising results, several limitations remain:

\begin{itemize}
\item \textbf{Single-center dataset:} The current dataset is sourced from a single institution. Although it exhibits variability in image quality, it may not cover all possible acquisition patterns or patient populations. Multi-center validation is needed to establish broader generalizability.

\item \textbf{Binary spine mask:} This study focuses on binary segmentation of the entire spine region. For detailed clinical tasks, vertebra-level segmentation or landmark detection may be required. Extending R2MF-Net to handle multi-class labels or joint segmentation and landmark regression is a possible direction~\cite{janssens2018fully}.

\item \textbf{2D modeling:} R2MF-Net operates on 2D radiographs independently. While this is appropriate for X-ray data, it does not exploit temporal sequence information (e.g., repeated examinations over time) or potential 3D reconstruction cues from biplanar systems~\cite{benameur2003three}.

\item \textbf{Model complexity:} Although computationally feasible, R2MF-Net is more complex than basic U-Net models. Deployment on resource-limited devices (e.g., embedded systems) may require further pruning or distillation.
\end{itemize}

\subsection{Future Work}
Future research may move along several directions:

\begin{itemize}
\item \textbf{Lightweight variants:} Designing parameter-efficient versions of R2MF-Net tailored to mobile devices or low-cost hardware, possibly through neural architecture search, pruning or knowledge distillation~\cite{chao2019hardnetseg}.

\item \textbf{Joint vertebra-level modeling:} Incorporating vertebral instance segmentation or keypoint detection within the same framework, enabling end-to-end estimation of clinical indices directly from X-rays.

\item \textbf{Multi-modal integration:} Combining radiographic images with clinical data (age, sex, Risser sign) or complementary imaging (e.g., low-dose CT) may further improve performance in atypical cases~\cite{trobisch2010idiopathicscoliosis}.

\item \textbf{Uncertainty modeling:} Incorporating probabilistic outputs or uncertainty estimation could help flag low-confidence cases for manual review, integrating the system more safely into clinical workflows.
\end{itemize}

\section{Conclusion}
This paper presents R2MF-Net, a recurrent residual multi-path segmentation network designed for robust multi-directional spine X-ray segmentation. The network integrates a cascade of coarse and fine segmentation stages with R2-Jump connections, MC-Skip multi-scale fusion and SCSE-Lite attention, enabling accurate delineation of spinal structures under diverse imaging conditions. Extensive experiments on a real clinical dataset demonstrate that R2MF-Net consistently outperforms strong CNN- and Transformer-based baselines in IoU, Dice and boundary accuracy~\cite{chen2018deeplabv3plus,oktay2018attentionunet,chao2019hardnetseg,chen2021transunet,hatamizadeh2022swinunet}.

By producing reliable spine masks in coronal and bending views, R2MF-Net provides a solid foundation for automated scoliosis assessment and has the potential to reduce radiologist workload and improve measurement reproducibility. Future work will explore lightweight variants, vertebra-level modeling and multi-center validation to further advance the deployment of such systems in daily clinical practice.

\bibliographystyle{unsrt}
\bibliography{R2MF}

\end{document}